\documentclass{article}

\usepackage[english]{babel}

\usepackage[letterpaper,top=2cm,bottom=2cm,left=3cm,right=3cm,marginparwidth=1.75cm]{geometry}

\usepackage{amsmath}
\usepackage{graphicx}
\usepackage[colorlinks=true, allcolors=blue]{hyperref}
\usepackage{amsthm}
\usepackage{tikz}
\usepackage{centernot}
\usepackage{mathtools}
\usepackage{ stmaryrd }
\usetikzlibrary{matrix,arrows.meta}
\usepackage{amsmath}
\usepackage{tikz}
\usetikzlibrary{matrix,arrows,decorations.pathmorphing}
\usetikzlibrary{shapes,arrows,positioning}
\usepackage{graphicx}
\usepackage{amsmath}
\usepackage{amssymb}
\usepackage{latexsym}
\usepackage{amssymb,amsmath}
\usepackage{tikz-cd}
\usetikzlibrary{arrows}
\usepackage{amsfonts}
\usepackage{amssymb}
\usepackage{graphicx}
\usepackage{fancyhdr}
\usepackage{fancyvrb}
\usepackage{fancybox}
\usepackage{float}
\usepackage{geometry}
\usepackage{minitoc}
\usepackage{indentfirst}
\usepackage{multicol, color}
\usepackage[outerbars]{changebar}
\usepackage{pifont,textcomp}
\usepackage{amsfonts}
\usepackage{amsmath}
\usepackage{amscd}
\usepackage{array}
\usepackage{multirow}
\usepackage{mathrsfs}
\usepackage{amssymb}
\usepackage{amsthm}
\usepackage{amsmath}
\usepackage{subfig}
\usepackage{graphicx}
\usepackage[colorlinks=true, allcolors=blue]{hyperref}

 \title{Gazeau-Klauder coherent states for a harmonic position-dependent mass}
 \author{  Daniel Sabi Takou$^{1,2}$, 	Assimiou Yarou Mora$^{2}$,  \\	Ibrahim Nonkan\'e$^{3}$, 	Lat\'evi M. Lawson$^{4,5}$   and  Gabriel Y. H. Avossevou$^{2}$
 	\space\\\\
 	$^{1}$Ecole Polytechnique d'Abomey Calavi (EPAC-UAC),\\
	Universit\'e d'Abomey-Calavi (UAC), B\'enin\\
	$^{2}$ Unit\'e de Recherche en Physique Th\'eorique (URPT),\\
	Institut de Math\'ematiques et de Sciences  Physiques (IMSP),\\
		01 B.P. 613 Porto-Novo, Rep. du B\'enin\\
			${}^3$Département d'\'Economie et de Math\'ematiques Appliqu\'ees,\\ IUFIC, Universit\'e Thomas Sankara, Burkina faso\\ 
	$^{4}$ African Institute for Mathematical Sciences (AIMS) Ghana, Accra, Ghana.\\
	$^{5}$Universit\'e de Lom\'e, Facult\'e des Sciences, D\'epartement de Physique, Lom\'e, Togo\\\\		 
 sabitakoudaniel11@gmail.com$^{1,2}$,  assimiouyaroumora@gmail.com$^{2}$,\\  inonkane@univouaga2.bf$^{3}$, latevi@aims.edu.gh$^{4,5}$ and gabriel.avossevou@imsp-uac.org$^{2}$ }

\date{}

\begin{document}
\maketitle

\begin{abstract}
In this paper, we study the dynamic of position-dependent mass system confined in harmonic oscillator potential. We derive the eigensystems by solving  the Schrödinger-like equation which describes this system.        We construct coherent states \`{a} la Gazeau-Klauder for this system. We show that these  states satisfy  the Klauder's mathematical condition to build coherent states. We compute and analyse some statistical properties of these states. We find that these states  exhibit sub-Poissonian statistics. We also evaluate quasiprobability distributions such as the Wigner function  to
demonstrate graphically nonclassical features of these states.
\end{abstract}
\section{Introduction}
The dynamic of systems with position-dependent mass (PDM) have found in the last decades applications in the  fields of material science \cite{1,2,3} and condensed matter physics \cite{4,5}. The majority of such studies dedicated to the problems relevance to semiconductor physics and solid state physics \cite{5a,5b,5c,5d}.
Recently, the study of a free PDM system in quantum gravitational background fields have been realized \cite{6a,6b,6c,6d}. Further, the asymmetric forms of PDM of type 
\begin{eqnarray}
	m(x)=\frac{m_0}{(1+\alpha x)^2},
\end{eqnarray}
have successfully been explained in \cite{6d,6d1,6e,6f}, where $m_0$ is the constant mass and $\alpha$ is a strength parameter that
measures the extent to which mass is depending on the position. In the present paper, we propose a quadratic form of this PDM of  type 
\begin{eqnarray}\label{z}
	m(x)=\frac{m_0}{(1+\alpha x^2)^2},
\end{eqnarray}
confined in a one dimensional (1D) harmonic oscillator potential. This model generalizes the  PDM of type introduced in \cite{6d,6e,6f,6g,6h,6i}.  In the current paper, we solve the Schr\"{o}dinger-like equation of this model \eqref{z}  in 1D harmonic  potential. The eigenvalues and the eigenfunctions of this equation is obtained, and we  observe that the energy levels are curved by the deformed parameter $\alpha$   and they    increase with this parameter.  With this result at hand,  we construct the Gazeau-Klauder (GK) coherent states \cite{7} for the spectrum of this position-deformed mass system. We show
that these states satisfy the Klauder’s mathematical conditions to build coherent states \cite{8,9}. We also explore the 
statistical properties  of these states, such as the photon distribution, the photon mean number, the
intensity correlation, the Mandel parameter and the Wigner. We find that these states are sub-Poissonian in nature and demonstrate nonclassical features.

The paper is organized as follows: In the next section, we study the dynamics of PDM in a harmonic oscillator potential. In Section \eqref{sec3}, we construct GK \cite{7}
coherent states for the deformed spectrum of PDM \eqref{z}. We discuss the quantum statistical properties of the constructed coherent states, and we show that theses states have sub-Poisson statistics.
Finally, we conclude this work in section \eqref{sec4}.

\section{Harmonic   position-dependent mass Schr\"{o}dinger equation}\label{sec2}
\begin{figure}  
  \centering
 \includegraphics[width=8cm, height=5cm]{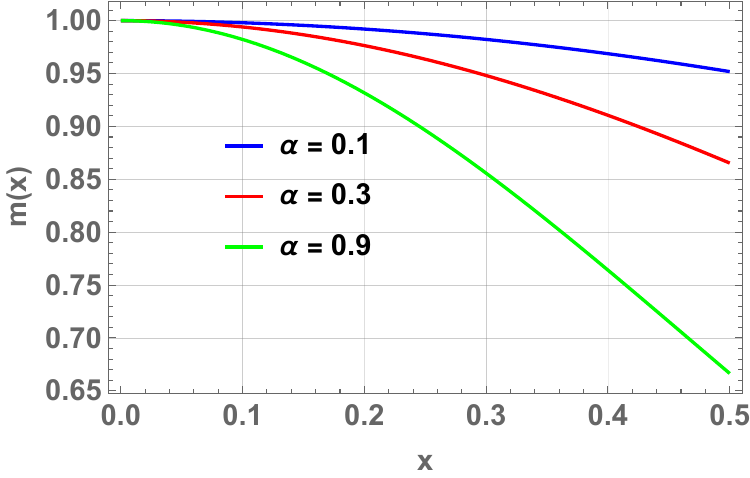}
   \caption{PDM versus the position $x$ for different values of $\alpha$.}
   \label{pdm1}
\end{figure}

The most general Hermitian kinetic Hamiltonians for a particle with position-dependent mass (PDM) $m(\hat x)$ can be writen in the form \cite{10}
\begin{eqnarray}
\hat T=\frac{1}{4}\left[m^\alpha(\hat x)\hat p m^\beta \hat pm^\gamma(\hat x)+ m^\gamma(\hat x)\hat p m^\beta \hat pm^\alpha(\hat x)\right],
\end{eqnarray}
where $\alpha, \beta$ and $\gamma$ are parameters satisfying the constraint   $\alpha +\beta+\gamma=-1$. Clearly, there
are different Hamiltonians depending on the choices of the parameters \cite{11,11a,11b,11c,11d}. Here we shall work
with the Mustafa and  Mazharimousavi form \cite{11d} which corresponds to the choice $\alpha=\gamma=-\frac{1}{4}$, and  $\beta=-\frac{1}{2}$  such as
\begin{eqnarray}
\hat T=\frac{1}{2} \frac{1}{m^{\frac{1}{4}}(\hat x)} \hat p{\frac{1}{m^{\frac{1}{2}}(\hat x)}} \hat p\frac{1}{m^{\frac{1}{4}}(\hat x)} (\hat x)  .
\end{eqnarray}
 The corresponding Hamiltonian of the system is given by
 
 \begin{eqnarray}
\hat H=\hat T+V=\frac{1}{2} \frac{1}{m^{\frac{1}{4}}(\hat x)} \hat p{\frac{1}{m^{\frac{1}{2}}(\hat x)}} \hat p\frac{1}{m^{\frac{1}{4}}(\hat x)} +V(\hat x), 
\end{eqnarray}
 where  $V(\hat x)$ is  the potential energy. The time-independent Schr\"{o}dinger equation  for  a harmonic oscillator particle, i.e. $ V( x)=\frac{1}{2}m_0\omega^2 x^2$  is given by 
\begin{eqnarray}\label{a}
E\phi(x)=-\frac{\hbar^2}{2m_0}\sqrt[4]{\frac{m_0}{m(x)}}\frac{d}{dx}\sqrt{\frac{m_0}{m( x)}}\frac{d}{dx}\sqrt[4]{\frac{m_0}{m( x)}}\phi(x)+V(x)\phi(x),
\end{eqnarray}
where $m_0$ is a constant mass,  $ E$         is the energy spectrum, $\phi(x)$ is the wavefunction defined on a Hilbert space $\mathcal{H}=\mathcal{L}^2(\mathbb{R})$. For the present case, we choose the PDM  $m(x)$ profile as
 \begin{eqnarray}\label{71}
  m(x)=\frac{m_0}{(1+\alpha x^2)^2},
 \end{eqnarray}
 where $\alpha$ is a deformed parametr $0<\alpha<1.$ This PDM generalised the ones introduced in \cite{6d,6e,6f,6g,6h,6i}. It could be identified as an inverse square length parameter related to a measure of curvatures or holes in the given physical system.
 The PDM is illustated in Fig.\ref{pdm1} as a function of the position $x$ $(0 < x < 0.5)$. In this description, the effective mass  decreases when we increase the parameter $\alpha$. This indicates that  the increasing of the given system curvatures lower the mass of particle.

\begin{figure}
 \centering
\label{key}		\includegraphics[width=8cm, height=6cm]{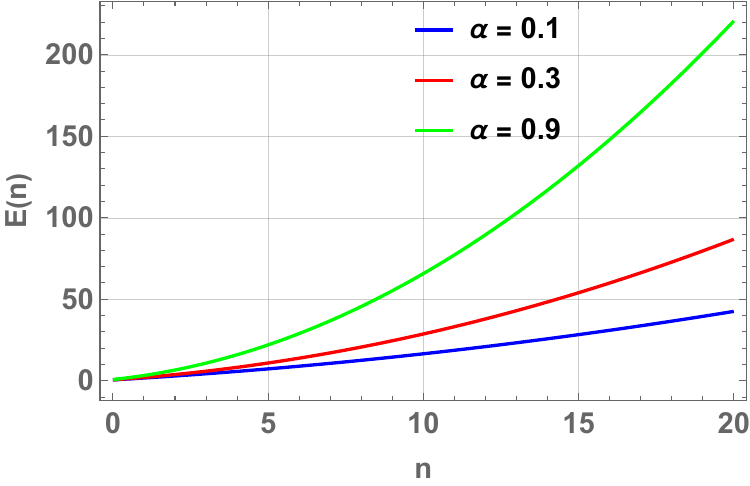}
    \caption{The energy for a PDM harmonic oscillator versus the quantum number $n$ for fix values of the parameter $\alpha$.
}\label{en}
\end{figure}

The Eq.\eqref{a} can be conveniently rewritten by means of the transformation $\phi(x)=\sqrt[4]{m(x)/m_0}\psi(x)$ as is 
 \begin{eqnarray}\label{schro}
  E\psi(x)=-\frac{\hbar^2}{2m_0}\left(\sqrt{\frac{m_0}{m(x)}}\frac{d}{dx}\right)^2 \psi(x)+ \frac{1}{2}m_0\omega^2 x^2 \psi(x).
 \end{eqnarray}
 The latter Schr\"{o}dinger equation can be rewritten as follows
  \begin{eqnarray}\label{schro}
  E\psi(x)&=&-\frac{\hbar^2}{2m_0}\left[(1+\alpha x^2) \frac{d}{d x}\right]^2 \psi(x)+ \frac{1}{2}m_0\omega^2 x^2 \psi(x).
 \end{eqnarray}
A change of variable from $x$ to
\begin{eqnarray}
   q= \rho\sqrt{\alpha}=\arctan(x\sqrt{\alpha}),
\end{eqnarray}
maps the region $-\infty < x <\infty$ to $-\frac{\pi}{2}< q <\frac{\pi}{2}.$ We simplify Eq.\eqref{schro} to
\begin{eqnarray}
    \frac{d^2\psi}{dq^2}+\left(\varepsilon-\kappa^2 \frac{s^2}{c^2} \right)\psi&=&0,
\end{eqnarray}
where 
\begin{eqnarray}
    \varepsilon=\frac{2m_0E}{\alpha\hbar^2},\quad
    \kappa=\frac{m_0\omega}{\alpha\hbar}\quad c=\cos q, \quad \mbox{and}\quad s=\sin q.
\end{eqnarray}
Let $\psi(q) = c^\lambda f(s)$ where $\lambda $ is a constant to be determined. Then the equation for $f(s)$ is
\begin{eqnarray}\label{k}
    (1-s^2)\frac{d^2 f}{ds^2}-(2\lambda+1)s\frac{d f}{ds}+\left((\varepsilon-\lambda)-\{\kappa^2-\lambda(\lambda-1)\}\frac{s^2}{c^2}\right)f=0.
\end{eqnarray}
The variable is now  $-1 < s < 1$.  We
fix $\lambda $ by requiring the coefficient of the tangent squared term to vanish:
\begin{eqnarray}
    \kappa^2-\lambda(\lambda-1)=0.
\end{eqnarray}
The wave function should be non–singular at $c = 0$, which implies
\begin{eqnarray}
    \lambda=\frac{1}{2}+\frac{1}{2}\sqrt{1+4\kappa^2}.
\end{eqnarray}
This simplifies Eq.\eqref{k} to
\begin{eqnarray}\label{l}
    (1-s^2)\frac{d^2 f}{ds^2}-(2\lambda+1)s\frac{d f}{ds}+(\varepsilon-\lambda)f=0.
\end{eqnarray}
Similarly, $f(s)$  should be non-singular at $s = \pm 1$. Thus we require a polynomial solution to Eq.\eqref{l}. This requirement imposes the following condition on the coefficient of $f$:
\begin{eqnarray}\label{e}
    \varepsilon-\lambda=n(n+2\lambda),
\end{eqnarray}
where $n$ is a non–negative integer \cite{13}. Equation \eqref{l} becomes
\begin{eqnarray}\label{eq3}
    (1-s^2)\frac{d^2 f}{ds^2}-(2\lambda+1)s\frac{d f}{ds}+n(n+2\lambda)f=0.
\end{eqnarray}
The energy eigenvalues follow from the condition Eq.\eqref{e}:
\begin{eqnarray}\label{en11}
    E_n=\hbar\omega \left(n+\frac{1}{2}\right)\sqrt{1+\frac{\alpha^2\hbar^2}{4m_0^2\omega^2}}+\frac{\alpha\hbar^2}{2m_0}\left(n^2+2n+\frac{1}{2}\right).
\end{eqnarray}
This spectrum is no longer different to the spectrum of a harmonic oscillator  in momentum-deformed Heisenberg algebra with minimal length uncertainty \cite{14}.
At the limit $\alpha\rightarrow 0$, we recover the spectrum of 1D ordinary harmonic oscillator such that
\begin{eqnarray}
    \lim_{\alpha\rightarrow 0} E_n=\hbar\omega \left(n+\frac{1}{2}\right).
\end{eqnarray}
Figure \eqref{en} illustrates the energy levels \eqref{en11}  
of the system as functions of the quantum number $n$ for different values of the deformation parameter $\alpha$. One observes that the energy $E_n$ increases with $\alpha$. Since the effective mass, $m(x)$, in our description Fig.\eqref{pdm1} decreases when  $\alpha$ increases, and the lower the mass, the bigger is the energy of the particle. 
 This indicates that, the deformed parameter $\alpha$ induces a more pronounced dilatation of quantum levels which, consequently implies the increase 
of energy band structures.

The solution for the differential equation \eqref{schro} becomes
\begin{eqnarray}
    \psi_n(q)=Nc^{\lambda} C_n^\lambda(q),
\end{eqnarray}
where $N$ is a normalization constant and $ C_n^\lambda(q)$  are the Gegenbauer polynomials \cite{13}.  Then by normalization $\langle \psi_n|\psi_n\rangle=1$, we have
\begin{equation}
    N^2 \int_{-1}^{1} (1 - s^2)^{\lambda - \frac{1}{2}} | C_n^\lambda(s)|^2 \, ds = 1.
\end{equation}
Using the following property of Gengenbauer’s polynomials \cite{13}
\begin{equation}
    N^2\int_{-1}^{1}du (1 - u^2)^{\beta - \frac{1}{2}} | C_n^\beta(u)|^2  =  \frac{\pi 2^{1 - 2\beta} \Gamma(n + 2\beta)}{n! (n + \beta) \Gamma^2(\beta)},
\end{equation}
we get for the normalized energy wave function the expression
\begin{equation}
    N = \sqrt{\frac{n! (n + \lambda) \Gamma^2(\lambda)}{\pi 2^{1 - 2\lambda} \Gamma(n + 2\lambda)}}.
\end{equation}
The final solution of  equation \eqref{a} becomes
\begin{eqnarray}
    \phi_n(q)&=&\frac{\psi(q)}{\sqrt{1+\tan^2q}}=N c^{\lambda+1} C_n^\lambda(s),\cr
     &=& \sqrt{\frac{n! (n + \lambda) \Gamma^2(\lambda)}{\pi 2^{1 - 2\lambda} \Gamma(n + 2\lambda)}}.c^{\lambda+1} C_n^\lambda(s),\label{di}
\end{eqnarray}
where 
\begin{eqnarray}
    c&=&\cos q=\frac{1}{\sqrt{1+\alpha x^2}},\\
    s&=&\sin q =\frac{x\sqrt{\alpha}}{\sqrt{1+\alpha x^2}}.
\end{eqnarray}

As we can  remark with this approach, the determination of the spectrum allowed
the introduction of the  discrete eigenbasis. Thus, to convert this spectrum
into  continuous spectrum, it is useful to introduce a continuous basis in which the diagonalization is possible. In this sense, the coherent states are the best candidates to achieve this purpose. In the literature, various coherent states  are contructed  \cite{15,16,17,18,19} and  have found considerable applications in different fields
of theoretical and experimental physics \cite{20,21,22,23,24,25,26,27,27a,28,28a}. However, in the present context, we construct  coherent states \`{a} la Gazeau-Klauder \cite{7} for this system. Based on the results \cite{6c,6g,15,28,30,31}, we show that   these  states satisfy  the Klauder's mathematical condition to build coherent states \cite{8,9} and we compute and analyse some statistical properties of these states.

\section{GK coherent states for a harmonic position-dependent mass }\label{sec3}
The GK-coherent states \cite{7} for a Hermitian Hamiltonian $\hat H$ with discrete, bounded below and nondegenerate eigenspectrum are defined as a two parameter set
\begin{equation}\label{coh}
|J,\gamma\rangle=\frac{1}{\mathcal{N}(J)}\sum_{n=0}^{\infty}\frac{J^{\frac{n}{2}}e^{-i\gamma e_n}}{\sqrt{\rho_n}}|\phi_n\rangle,
\end{equation}
where $J\in \mathbb{R}^+$,\,\,$\gamma\in\mathbb{R}$. The states $|\phi_n\rangle$ are the  orthogonal eigenstates of $\hat H$, such 
that is $\hat H |\phi_n\rangle=  e_n|\phi_n\rangle$ with
\begin{eqnarray}
    e_n=E_n-E_0 \quad \mbox{with}\quad  0=e_0<e_1<e_2 \cdots,
\end{eqnarray}
the  normalization constant $\mathcal{N}(J)$ is  chosen so that
\begin{eqnarray}\label{nor}
\langle \gamma, J|J,\gamma\rangle=
\mathcal{N}^{-2}(J)\sum_{n=0}^\infty\frac{J^n}{\rho_n}=1\implies \mathcal{N}^2(J)=\sum_{n=0}^\infty \frac{J^n}{\rho_n}.
\end{eqnarray}
The allowed values of $J$, $0 <J <R$ are determined by the radius of convergence $R =\lim_{n \rightarrow \infty } (\rho)^{1/n}$ in the series defining $\mathcal{N}^2(J)$. The  moments of a probability
distribution $\rho_n$ is given by
\begin{eqnarray}\label{rho}
\rho_n=\int_{0}^{R}x^n\rho(x)dx=\prod_{k=1}^n e_k,\quad \rho_0=1.
\end{eqnarray}
We recover from the coherent states (\ref{coh}), the usual canonical coherent states by setting  $z=\sqrt{J} e^{-i\gamma}$ and $\mathcal{N}^{-1}(J)=e^J$ \cite{7} given by
\begin{eqnarray}\label{ord}
 |z\rangle = e^{-\frac{1}{2}|z|^2}\sum_{n=0}^{\infty}\frac{z^n}{\sqrt{n}!}|\phi_n\rangle.
\end{eqnarray}
The GK-coherent states (\ref{coh})  have to satisfy the following properties \cite{7,8,9}:
\begin{enumerate}
 \item Normalizability: $ \langle \gamma, J|J,\gamma\rangle=1.   $
 \item  Continuity: the mapping $|J',\gamma'\rangle\rightarrow$ $|J,\gamma\rangle$ is continuous in some appropriate topology.
 \item Resolution of unity: $\int |J,\gamma \rangle\langle \gamma, J| d\mu(J,\gamma)=\mathbb{I}$.
 \item  Temporal stability: $ e^{-i H t}|J,\gamma,\rangle=|J,\gamma +\nu t\rangle$,\,\, with \,\,\,$\nu$ = const.
\end{enumerate}

\subsection{ Construction of GK coherent states}
Let us recall the  energy eigenvalues defined in equation (\ref{en1}) as
\begin{eqnarray}\label{en1}
    E_n=\hbar\omega \left(n+\frac{1}{2}\right)\sqrt{1+\frac{\alpha^2\hbar^2}{4m_0^2\omega^2}}+\frac{\alpha\hbar^2}{2m_0}\left(n^2+2n+\frac{1}{2}\right)\quad \mbox{and}\quad E_0=\frac{1}{2}\hbar\omega \sqrt{1+\frac{\alpha^2\hbar^2}{4m_0^2\omega^2}}+\frac{\alpha\hbar^2}{4m_0}.
\end{eqnarray}
For the system under consideration, the dimensionless form of the latter energy is given by
\begin{eqnarray}
    e_n&=&E_n-E_0=n\hbar\omega \sqrt{1+\frac{\alpha^2\hbar^2}{4m_0^2\omega^2}}+\frac{\alpha\hbar^2}{2m_0}\left(n^2+2n\right)= \left(\hbar\omega \sqrt{1+\frac{\alpha^2\hbar^2}{4m_0^2\omega^2}}+\frac{\alpha\hbar^2}{m_0}\right)n+\frac{\alpha\hbar^2}{2m_0}n^2\cr
    &=&an^2+bn,\label{q1}
\end{eqnarray}
where the constants $a$ and $b$ are given by
\begin{eqnarray}
    a=\frac{\alpha\hbar^2}{2m_0\omega},\quad \mbox{and}\quad b=\left(\hbar \omega\sqrt{1+\frac{\alpha^2\hbar^2}{4m_0^2\omega^2}}+\frac{\alpha\hbar^2}{m_0\omega}\right).
\end{eqnarray}
  The dimensionless form of the  energy \eqref{q1} is similar to the ones obtained in \cite{15,30} used to study the laser light propagation in a nonlinear Kerr medium.
The product of
these dimensionless energies $e_n$  represented by  $\rho_n$ is defined as
\begin{eqnarray}
\rho_n &=& \prod_{k=1}^n e_k, \quad\mbox{with}\quad e_i = a k^2 + b k\cr
        &=& \prod_{k=1}^n k  \prod_{k=1}^n (a k + b).\label{q1}
\end{eqnarray}
With the following computations
\begin{eqnarray}
\prod_{k=1}^n k = n!=\Gamma(n+1), \quad \prod_{k=1}^n (a k + b) = a^n \prod_{k=1}^n \left( k + \frac{b}{a} \right)=a^n\frac{\Gamma\left( n + 1 + \frac{b}{a} \right)}{\Gamma\left( 1 + \frac{b}{a} \right)}.
\end{eqnarray}
The  equation \eqref{q1} becomes 
\begin{eqnarray}
\rho_n = n! a^n  \frac{\Gamma\left( n + 1 + \frac{b}{a} \right)}{\Gamma\left( 1 + \frac{b}{a} \right)}= a^n  \frac{\Gamma\left( n + 1  \right)\Gamma\left( n + 1 + \frac{b}{a} \right)}{\Gamma\left( 1 + \frac{b}{a} \right)},\quad \rho_0=1.
\end{eqnarray}
The normalization constant \eqref{nor} is calculated as
\begin{eqnarray}
N^2(J) = \Gamma\left( 1 + \frac{b}{a} \right) {}_0F_1\left(1 + \frac{b}{a}; \frac{J}{a} \right),
\end{eqnarray}
where $ {}_0F_1$ is  the hypergeometric function \cite{13},  and the radius of convergence turns out to be
\begin{eqnarray}
    R=\lim_{n\rightarrow\infty}[\rho_n]^{\frac{1}{n}}=\lim_{n\rightarrow\infty}\left[a^n \frac{\Gamma\left( n + 1 \right)\Gamma\left( n + 1 + \frac{b}{a} \right)}{\Gamma\left( 1 + \frac{b}{a} \right)}\right]^{\frac{1}{n}}=\infty.
\end{eqnarray}
As a result, the coherent states given in equation \eqref{coh} may be expressed as
\begin{eqnarray}\label{coh}
|J,\gamma\rangle&=&\frac{1}{\mathcal{N}(J)}\sum_{n=0}^{\infty}\frac{J^{\frac{n}{2}}e^{-i\gamma e_n}}{\sqrt{\rho_n}}|\phi_n\rangle\cr
&=&  \frac{1}{\sqrt{\Gamma\left( 1 + \frac{b}{a} \right) {}_0F_1\left( 1 + \frac{b}{a}; \frac{J}{a} \right)}} \sum_{n=0}^{\infty} \frac{J^{\frac{n}{2}} e^{-i \gamma (a n^2 + b n)}}{\sqrt{a^n \frac{\Gamma\left( n+1 \right)\Gamma\left( n + 1 + \frac{b}{a} \right)}{\Gamma\left( 1 + \frac{b}{a} \right)}}} |\phi_n(q)\rangle.\label{ch}
\end{eqnarray}
By multiplying the above equation by the vector $\langle q|$ we express the coherent states \eqref{ch}, in term of the discrete
wave function \eqref{di}
\begin{eqnarray}
    \phi_n(q,J,\gamma)&=&\frac{1}{\mathcal{N}(J)}\sum_{n=0}^{\infty}\frac{J^{\frac{n}{2}}e^{-i\gamma e_n}}{\sqrt{\rho_n}}\phi_n(q),\cr
     &=& \frac{1}{\mathcal{N}(J)}\sqrt{\frac{n! (n + \lambda) \Gamma^2(\lambda)}{\pi 2^{1 - 2\lambda} \Gamma(n + 2\lambda)}} \sum_{n=0}^{\infty}\frac{J^{\frac{n}{2}}e^{-i\gamma e_n}}{\sqrt{\rho_n}}  c^{\lambda+1} C_n^\lambda(s).
\end{eqnarray}

\subsection{Mathematical properties}
In this subsection, we will discuss the above properties of these states \eqref{ch} by analysing the non-orthogonality, normalizability,
the conditions of continuity in the label,  the resolution of identity by finding a positif  weight
function $\mathcal{W}(J)$ and the temporal stability.

\subsubsection{The non-orthogonality}

The overlap of two GK coherent states is given by

\begin{eqnarray}
	\langle J',\gamma'	|J,\gamma\rangle=	\frac{1}{\mathcal{N}(J')\mathcal{N}(J)}\sum_{n=0}^{\infty}\frac{\Gamma(1+\frac{b}{a})e^{-i(\gamma-\gamma') e_n}}{n!\Gamma(n+1+\frac{b}{a})}\frac{(JJ')^{\frac{n}{2}}}{a^n}.
\end{eqnarray}
This shows that the scalar product of two coherent states does not vanish.
However, for $J'=J$ and $\gamma'=\gamma$, the above relation provides us  the normalization condition $ \langle J,\gamma	|J,\gamma\rangle=1.$

\subsubsection{The Label continuity}
The label continuity condition of the $ |J,\gamma\rangle$ can then be stated as,
\begin{eqnarray}
	|||J'\gamma'\rangle - |J,\gamma\rangle ||^2=2[1-\mathcal{R}e(\langle J',\gamma	|J',\gamma'\rangle)]\rightarrow 0,\quad \mbox{when}\quad (J',\gamma')\rightarrow (J,\gamma).
\end{eqnarray}

\subsubsection{Resolution of unity}
\begin{figure}	
 \centering
\label{key}		\includegraphics[width=8cm, height=7cm]{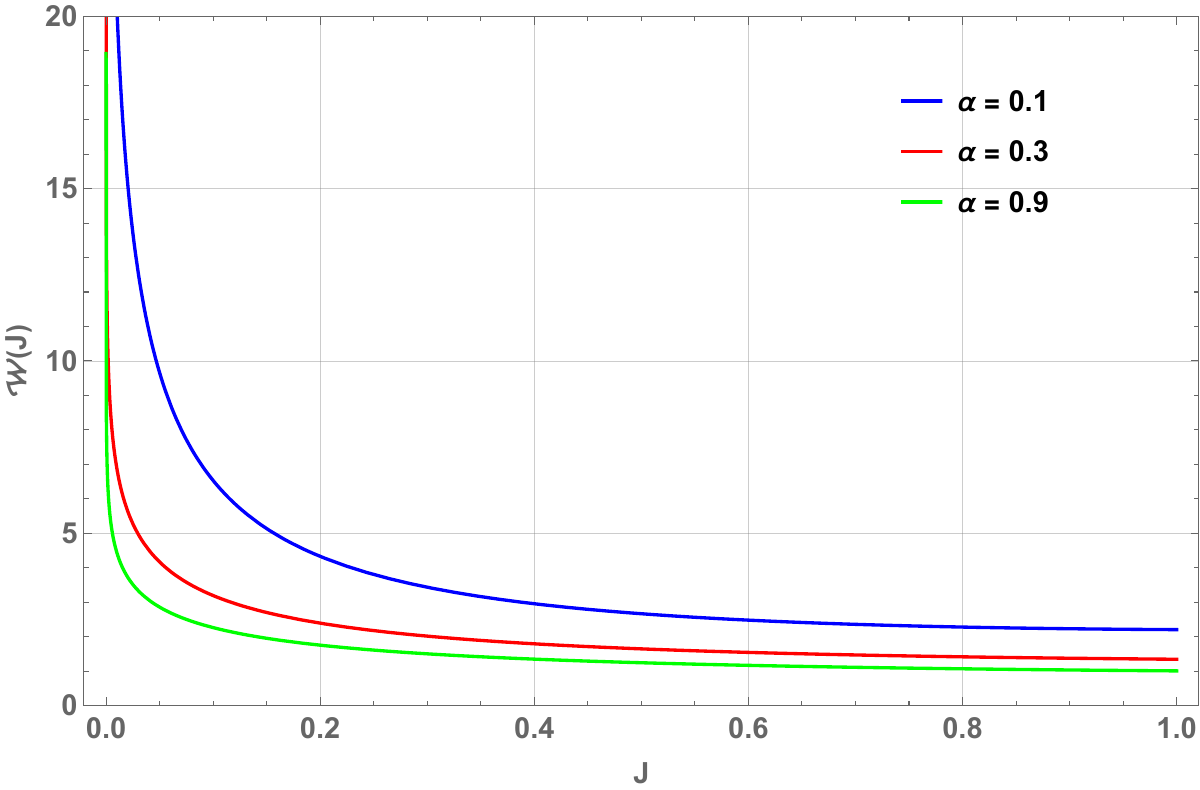}\caption{Weight function \eqref{weig}  versus the parameter $J$ for different values of $\alpha$.
}\label{en5}
\end{figure}

The overcompleteness relation reads as follows
\begin{eqnarray}\label{id}
	\int d\mu(J,\gamma)   |J,\gamma\rangle\langle J,\gamma|= \mathbb{I},
\end{eqnarray}
here the meseare $d\mu(J,\gamma)=\mathcal{W}(J) \frac{d Jd\gamma}{2\pi}$ \cite{6g}. 
By substituting equations (\ref{coh})   into equation (\ref{id}), we
obtain
\begin{eqnarray}
	\int_{0}^{\infty}\overline {\mathcal{W}}(J) J^nd J=  a^n  \frac{\Gamma\left( n + 1  \right)\Gamma\left( n + 1 + \frac{b}{a} \right)}{\Gamma\left( 1 + \frac{b}{a} \right)},
\end{eqnarray}
where $\overline {\mathcal{W}}(J)=\mathcal{W}(J)/ \mathcal{N}^2(J)$ and $\sum_{n=0}^\infty|\phi_n\rangle\langle \phi_n|=\mathbb{I}$. Perfoming $n=s-1$,   the integral from the above equation is called
the Mellin transform
\begin{eqnarray}\label{id2}
	\int_{0}^{\infty}\overline {\mathcal{W}}(J) J^{s-1}d J=a^{s-1}\frac{\Gamma\left( s  \right)\Gamma\left( s + \frac{b}{a} \right)}{\Gamma\left( 1 + \frac{b}{a} \right)}.
\end{eqnarray}
Using the definition of Meijers G-function, it follows that 
\begin{eqnarray}\label{eqm} 
	\int_0^\infty dx x^{s-1}G_{p,q}^{m,n}
    	\left(\tau x\big|_{d_1,\cdots,\quad d_m,\quad d_{m+1},\quad \cdots, d_q}^{c_1,\cdots,\quad c_n,\quad c_{n+1},\quad \cdots, c_p}\right)\cr=
	\frac{1}{\tau^s}
\frac{\prod_{j=1}^m\Gamma(d_j+s)\prod_{j=1}^n\Gamma(1-c_j-s)}{\prod_{j=m+1}^q\Gamma(1-d_j-s)\prod_{j=n+1}^p\Gamma(c_j+s)}.
\end{eqnarray}
Comparing equations (\ref{id2}) and (\ref{eqm}), we obtain that
\begin{eqnarray}
\overline {\mathcal{W}}(J)=\frac{1}{a\Gamma(1+\frac{b}{a})}G_{0,0}^{2,0}
\left(\frac{J}{a} \Bigg|_{0,\,\,\frac{b}{a}}^{.\,.\,.}\right).
\end{eqnarray}
Since $\overline {\mathcal{W}}(J)=\mathcal{W}(J)/ \mathcal{N}^2(J)$, we finally get 
\begin{eqnarray}\label{weig}
{\mathcal{W}}(J)=\frac{ {}_0F_1\left(  1 + \frac{b}{a}; \frac{J}{a} \right)}{a}    G_{0,0}^{2,0}
\left(\frac{J}{a} \Bigg|_{0,\,\,\frac{b}{a}}^{.\,.\,.}\right).
\end{eqnarray}
Figure\eqref{en5} illustrates the weight function \eqref{weig} versus $J$ for various values of the deformed parameter $\alpha$. One can
observe that, the weight function globally decreases when the parameter $\alpha$ increases.
\subsubsection{The temporal stability}
The time evolution of coherent states $|J,\gamma\rangle$ can be obtained by unitary transformation $|J,\gamma,t\rangle=\hat U(t)|J,\gamma\rangle$ where the time evolution operator is given as $\hat U(t)=e^{-i\nu e_n t}$. In the present case, the time evolution of the GK coherent states \eqref{coh} is
given by
\begin{eqnarray}
	|J,\gamma,t\rangle= \frac{1}{\mathcal{N}(J)}\sum_{n=0}^{\infty}\frac{J^{\frac{n}{2}}e^{-i e_n (\gamma+\nu t)}}{\sqrt{\rho_n}}|\phi_n\rangle=|J,\gamma+\nu t\rangle.
\end{eqnarray}
By multiplying the above equation by  the vector  $\langle q|$, we have
\begin{eqnarray}
	\phi_n(q,J,\gamma,t)&=& \frac{1}{\mathcal{N}(J)}\sum_{n=0}^{\infty}\frac{J^{\frac{n}{2}}e^{-i e_n (\gamma+\nu t)}}{\sqrt{\rho_n}}\phi_n(q).
\end{eqnarray} 

\subsection{ The statistical properties}
After mathematical construction of the GKCSs, in the present subsection, we
investigate some of the quantum statistical properties of these states, such as the photon-number distribution and  the nonclassicality features decribed by  the Mandel parameter or the intensity correlation function  and the quasiprobability distribution function

\subsubsection{Photon-number distribution  }
\begin{figure}	
 \centering
\label{key}		\includegraphics[width=6cm, height=5cm]{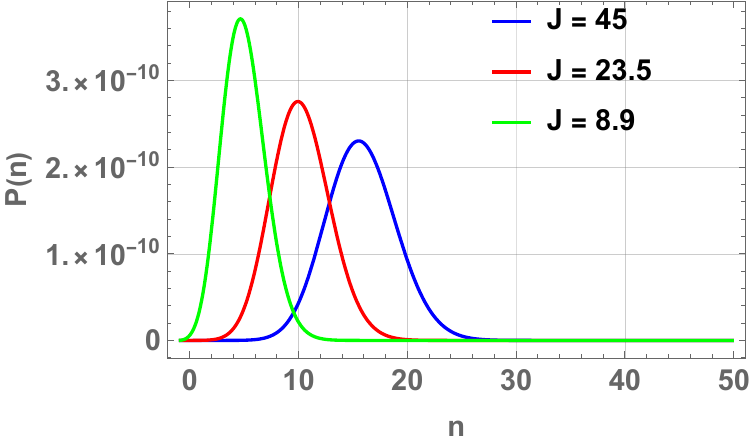}
\label{key}		\includegraphics[width=6cm, height=5cm]{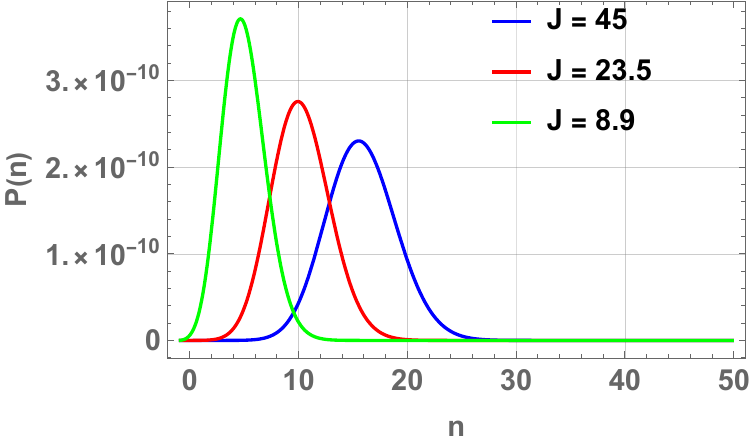}
\label{key}		\includegraphics[width=6cm, height=5cm]{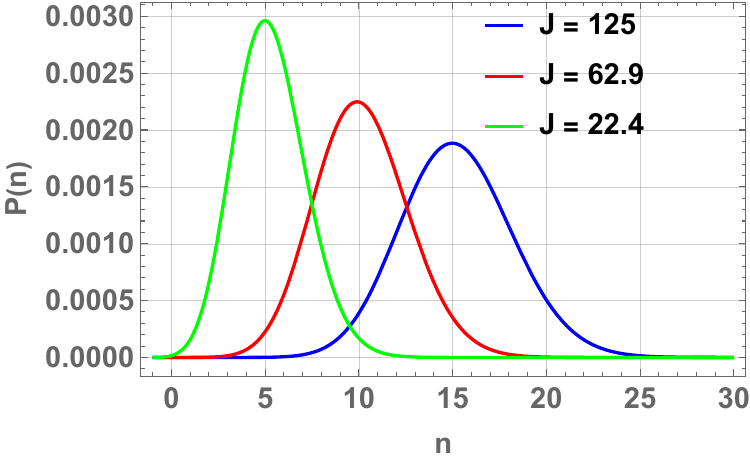}
\label{key}		\includegraphics[width=6cm, height=5cm]{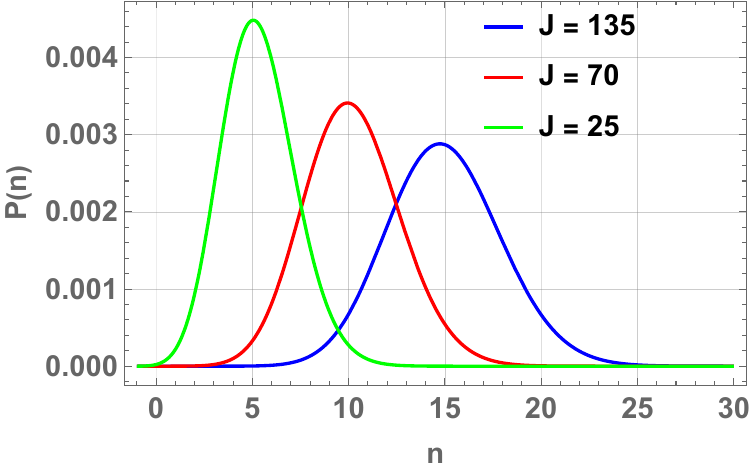}
\caption{The probability distribution \eqref{7} as a function of quantum number $n$ for different values of the coherent state parameter $J$ and the deformed parameters $\alpha=0.2$, $\alpha=0.3$ for the first two figures, respectively,
 and $\alpha=0.8$, $\alpha=0.9$ for the second two figures below, respectively.
}\label{7}
\end{figure}

The probability of finding the $n^{th}$ photons in the states $	|J,\gamma\rangle$ is given by
\begin{eqnarray}\label{p}
	P_n &=& |\langle\phi_n	|J,\gamma\rangle|^2=\frac{J^n}{\mathcal{N}^2(J)\rho_n}\cr
    &=& \frac{(J/a)^n}{ {}_0F_1\left(1 + \frac{b}{a}; \frac{J}{a} \right)\Gamma(n+1)\Gamma(n+1+\frac{b}{a})}.
\end{eqnarray}
Figure \eqref{7} illustrates the probability distribution plotted as a function of  the photon number $n$ for various values of the coherent state parameter $J$ and the deformed parameter $\alpha$. In these graphs, the coherent state parameter $J$ is selected so that the associated coherent states peak at the mean excitation quantum numbers $n = 5, 10,$ and $15$.

\subsubsection{Nonclassical properties}
Nonclassical properties of quantum states refer to characteristics that cannot be explained by classical physics. These properties often manifest in quantum systems by  defining the Mandel parameter  or the intensity correlation function or the Wigner function of GKCS \eqref{ch}. 

\begin{itemize}
    \item  The intensity correlation function or equivalently the Mandel $Q$-parameter yields the information about photon statistics of the quantum states. The intensity correlation function of the states \eqref{ch} is defined by 
\begin{eqnarray}
	g^{(2)}(0)=\frac{\langle  \hat N^2\rangle- \langle \hat  N\rangle }{[\langle \hat  N \rangle ]^2},
\end{eqnarray}
where $N$ is the number operator which is defined as the operator which diagonalizes the basis for the number states
\begin{eqnarray}
	\hat N |\phi_n\rangle=n |\phi_n\rangle\quad \mbox{and} \quad \hat N^2 |\phi_n\rangle=n^2 |\phi_n\rangle.
\end{eqnarray}
The Mandel $Q$-parameter is related to the intensity correlation function by 
\begin{eqnarray}
	Q= \langle  \hat N\rangle [ g^{(2)}(0)-1].
\end{eqnarray}
The intensity correlation function (or the Mandel $Q$-parameter) determines
whether the GKCSs  have a photon number distribution. This latter is sub-Poissonian if  $ g^{(2)}(0)<1$ (or $-1 \leqslant Q < 0$), Poissonian if $g^{(2)}(0)= 1$ or $( Q=0)$, and super-Poissonian if $g^{(2)}(0)>1$ (or $Q>0)$. 
We check that, for GKCSs (\ref{ch}), the expectation values of $\hat N$ and $\hat N^2$ can be computed as
\begin{eqnarray}
\langle  \hat N\rangle &=& \langle J,\gamma| \hat N |J,\gamma \rangle= \sum_{n=0}^\infty   n P_n= \left(\frac{J}{a+b} \right)
    \frac{{}_0F_1\left(2 + \frac{b}{a}; \frac{J}{a}\right)}
    {{}_0F_1\left(1 + \frac{b}{a}; \frac{J}{a}\right)},\\
\langle  \hat N^2\rangle&=&	 \langle J,\gamma| \hat N^2 |J,\gamma \rangle= \sum_{n=0}^\infty   n^2 P_n =  \left(\frac{J^2}{(a+b)(2a+b)} \right)\cr&&\times
    \frac{{}_0F_1\left(3+ \frac{b}{a}; \frac{J}{a}\right)}
    {{}_0F_1\left(1 + \frac{b}{a}; \frac{J}{a}\right)}+\left(\frac{J}{a+b} \right)
    \frac{{}_0F_1\left(2 + \frac{b}{a}; \frac{J}{a}\right)}
    {{}_0F_1\left(1 + \frac{b}{a}; \frac{J}{a}\right)}.
\end{eqnarray}
Taking into account the  latter results  of the expectation values of the number operator and its square, one gets
\begin{eqnarray}
	g^{(2)}(0)&=&\left(\frac{a+b}{2a+b}\right)\frac{{}_0F_1\left(1 + \frac{b}{a}; \frac{J}{a}\right){}_0F_1\left(3 + \frac{b}{a}; \frac{J}{a}\right)}{\left[{}_0F_1\left(2 + \frac{b}{a}; \frac{J}{a}\right)\right]^2},\\
Q &=&
\left(\frac{J}{2a+b}\right)\frac{{}_0F_1\left(3 + \frac{b}{a}; \frac{J}{a}\right)}{{}_0F_1\left(2 + \frac{b}{a}; \frac{J}{a}\right)}-\left(\frac{J}{a+b}\right)\frac{{}_0F_1\left(2 + \frac{b}{a}; \frac{J}{a}\right)}{{}_0F_1\left(1 + \frac{b}{a}; \frac{J}{a}\right)}.
\end{eqnarray}
In Figure \eqref{6}, the intensity correlation function $g^2
(0)$ and the Mandel $Q$-parameter have been plotted in terms of
the parameter $J$ for different values of the deformed parameter $\alpha$. One can see that the Mandel $Q$-parameter is
negative $Q<0$ and the intensity correlation function $g^2
(0)< 1$ which indicates that, the GKCS \eqref{ch} have sub-Poissonian statistics.

\begin{figure}	
 \centering
\label{key}		\includegraphics[width=7cm, height=6cm]{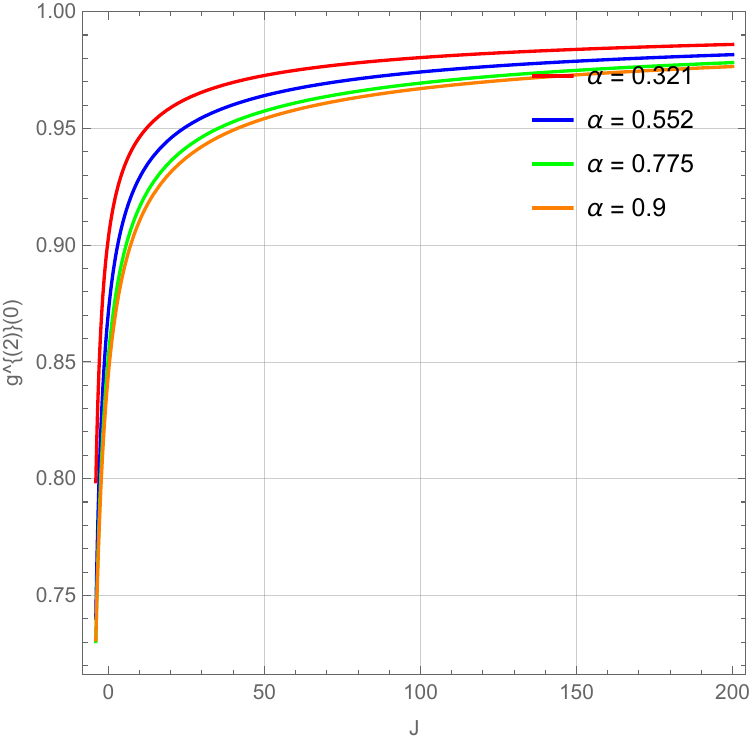}\quad
\label{key}		\includegraphics[width=7cm, height=6cm]{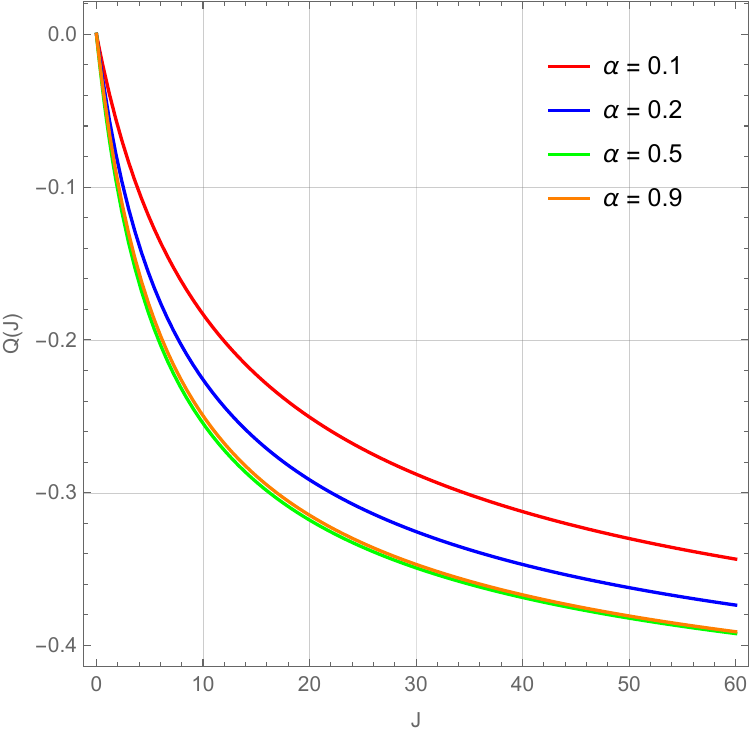}
\caption{The intensity correlation function $g^2(0)$
and Madel $Q$-parameter versus the corehent states parameter $J$ for different values of $\alpha$
}\label{6}
\end{figure}

\item The quasiprobability distribution function namely, the Wigner function  are given by \cite{15,32,33,34}
\begin{eqnarray}\label{w1}
    W(z)=\frac{2}{\pi}\sum_{n=0}^\infty(-1)^{n}\langle z|\hat \rho|z\rangle,
\end{eqnarray}
where $\hat \rho=|J,\gamma\rangle \langle J,\gamma|$ is the density matrix of GKCSs and $ |z\rangle $ is the ordinbary coherent states \eqref{ord}. The explicit form of equation \eqref{w1} reads
\begin{eqnarray}\label{w}
    W(z)= \frac{2}{\pi}\sum_{n=0}^\infty(-1)^{n}\left|\langle z|J,\gamma\rangle \right|^2 = \frac{2}{\pi}\frac{e^{-|z|^2}}{ {}_0F_1\left(1 + \frac{b}{a}; \frac{J}{a} \right)}\sum_{n=0}^\infty(-1)^{n}\left|\frac{J^{\frac{n}{2}}e^{-i\gamma e_n} z^n}{\sqrt{a^nn!\Gamma\left( n + 1  \right)\Gamma\left( n + 1 + \frac{b}{a} \right)}}\right|^2.
\end{eqnarray}
In figure \eqref{en15}, we  plot the Wigner function \eqref{w} for $J = 1$ and $\alpha= 0.1$. It is seen that the Wigner function
indeed takes negative values in some regions. This confirms nonclassical nature of the GKCSs the system.
\begin{figure}	
 \centering
\label{key}		\includegraphics[width=9cm, height=5cm]{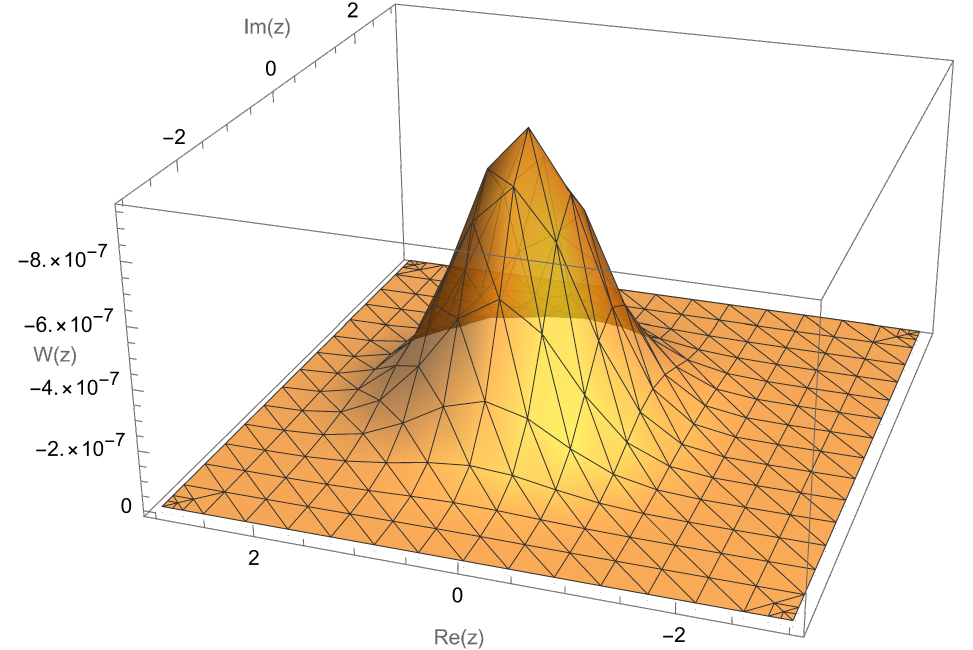}\caption{Weight function\eqref{w} for $J = 1$ and for $\alpha = 0.1$.
}\label{en15}
\end{figure}

\end{itemize}

\section{Conclusion}\label{sec4}
The Gazeau-Klauder coherent states for a position-dependent mass system (7) in a harmonic potential were built in the first section of this work. By determining its eigensystems, we were able to solve this system's Schrödinger-like   equation.
  We observed that the quantum energy $E_n$ levels are curved and are increased  with the deformed parameter $\alpha$.  In the second section of the paper, we created coherent states for these discrete eigensystems    \`{a} la Gazeau-Klauder. We have demonstrated that these states
We have demonstrated that these states meet the label continuity, normalizability, overcompleteness, and spatiotemporal stability requirements of Klauder's mathematical condition for building coherent states. We look at and analyze statistical features such as the Mandel parameter, the second-order correlation function, and the photon number distribution. We discovered that the statistics of these states were sub-Poissonian. Additionally, we have demonstrated statistically that these states display a negative Wigner function component, indicating their non-classicality.

\section*{Acknowledgments}
LML acknowledges support from AIMS-RIC Grant

\end{document}